\documentclass
[preprint,prb,a4paper,amsfonts,amssymb,floats,titlepage,fleqn,footinbib,12pt,onecolumn,nobalancelastpage]{revtex4}%
\usepackage{amsfonts}
\usepackage{amsmath}
\usepackage{amssymb}
\usepackage{graphicx}
\usepackage{longtable}%
\begin{document}
\author{M. ElMassalami, R. Moreno, R. M. Saeed, F. A. B. Chaves, C. M. Chaves}
\affiliation{Instituto de Fisica, Universidade Federal do Rio de Janeiro, Caixa Postal
68528, 21945-970 Rio de Janeiro, Brazil}
\author{H. Takeya}
\affiliation{National Institute for Materials Science,1-2-1 Sengen, Tsukuba, Ibaraki
305-0047, Japan}
\author{B. Ouladdiaf}
\affiliation{Institut Laue-Langevin, B.P.156 ,38042 Grenoble Cedex 9, France}
\author{M. Amara}
\affiliation{Institut N\'{e}el - CNRS, B\^{a}t. D, B.P. 166, 38042 Grenoble Cedex 9, France }
\title{On the ferromagnetic structure of the intermetallic borocarbide TbCo$_{2}%
$B$_{2}$C}
\date{\today{}}

\begin{abstract}
Based on magnetization, specific heat, magnetostriction, and neutron
diffraction studies on single-crystal \textrm{TbCo}$_{2}$\textrm{B}$_{2}%
$\textrm{C}, it is found out that the paramagnetic properties, down to liquid
nitrogen temperatures, are well described by a Curie-Weiss behavior of the
Tb$^{3+}$ moments. Furthermore, below $T_{\text{c}}$= 6.3 K, the Tb-sublattice
undergoes a ferromagnetic (FM) phase transition with the easy axis being along
the $(100)$ direction and, concomitantly, the unit cell undergoes a
tetragonal-to-orthorhhombic distortion. For fields up to 90 kOe, no
field-induced splitting of the Co 3\textit{d} orbitals was observed; as such
the internal field must be well below the critical value needed to polarize
the Co 3\textit{d} subsystem. The manifestation of a FM state in
\textrm{TbCo}$_{2}$\textrm{B}$_{2}$\textrm{C} is unique among all other
isomorphous borocarbides, in particular \textrm{TbNi}$_{2}$\textrm{B}$_{2}%
$\textrm{C }($T_{N}$=15 K, incommensurate modulated magnetic state) even
though the Tb-ions in both isomorphs have almost the same crystalline electric
field properties. The difference in the magnetic modes of these Tb-based
isomorphs is attributed to a difference in  their exchange couplings caused by
a variation in their lattice parameters and in the position of their Fermi levels.

\end{abstract}

\maketitle

\section{Introduction}

Whenever a family of compounds, containing 3\textit{d} transition-metal ($M$)
and rare-earth ($R$) atoms, manifests similar band structures, the trend in
their magnetic properties\ can be rationalized in terms of the position of the
Fermi level $E_{F}$ within the density of states $N$($E$) curve. The cubic,
Laves-type $RM_{2}$ family of compounds provides a best
illustration.\cite{Bloch-Lemaire70-RCo2,Bloch75-RCo2,Cyrot79-electr-cal,Cyrot79-b-electr-cal}%
\ Here, it is commonly considered that the $R$-ion moments are localized while
the $M$ moments, if they exist, are itinerant. Assuming further that the
mutual interactions among the magnetic ions can be described within the
molecular field theory and, in addition, using Landau phase-transition
arguments together with Stoner and Wohlfarth-Rhodes criteria, then the
magnetic properties of the whole $RM_{2}$ family can be classified according
to the first few energy derivatives of $N$($E$) evaluated at $E_{F}%
$%
.\cite{Bloch-Lemaire70-RCo2,Bloch75-RCo2,Cyrot79-electr-cal,Cyrot79-b-electr-cal}
This model is successful in explaining the evolution of the paramagnetic
susceptibility, the type of the magnetic phase transition, the relative
magnitude of the critical temperature, and, moreover, the magnetism of the
3\textit{d} subsystem; in particular it explains why the Ni sublattice is
magnetically silent in $R$\textrm{Ni}$_{2}$ while the Co subsystem has an
intermediate character: being strongly exchange-enhanced in $R$\textrm{Co}%
$_{2}$ ($R$ = Y, Lu) while developing a delocalized moment in magnetic
$R$\textrm{Co}$_{2}$. For such $R$\textrm{Co}$_{2}$ series, it was found out
that an adequate effective magnetic field at the Co site, $H_{\text{eff}%
}^{\text{Co}},$ is able, at and above a critical field $H_{\text{cr}%
}^{\text{Co}}$, to induce a metamagnetic transition due to which the Co atom
acquires an itinerant moment of $\sim$1$\mu_{\text{\textrm{B}}}$.

The above-mentioned interplay between the magnetism of $M$ and $R$ subsystems
is expected to be manifested also in the quaternary borocarbides
$RM_{\text{\textrm{2}}}$\textrm{B}$_{\text{\textrm{2}}}$\textrm{C} family (see
e.g. Ref. 5\nocite{Muller01-interplay-review} and references therein). Though
the investigation of this interplay is of interest for the understanding of
the magnetism (and superconductivity) of this family, there are only very few
studies bearing on this subject and these are mostly limited to the
$R$\textrm{Ni}$_{\text{\textrm{2}}}$\textrm{B}$_{\text{\textrm{2}}}$\textrm{C}
series. Nevertheless, the extensive magnetic characterization of this
$R$\textrm{Ni}$_{\text{\textrm{2}}}$\textrm{B}$_{\text{\textrm{2}}}$\textrm{C}
series did clarify the influence of $R$ ($M$) on their overall
magnetic\ (superconducting) properties;\cite{Muller01-interplay-review} in
particular, it is established that while the $M$-doping hardly influences the
magnetic properties (but leads invariably to a degradation of the
superconductivity), the variation in $R$ strongly modifies the magnetic
structure: for \textit{R }= Pr, Ho, Dy, the magnetic structures consist of a
commensurate antiferromagnetic, AFM,
state,\cite{Lynn97-RNi2B2C-ND-mag-crys-structure,Campbell-Ho-HT-Diagram,Dervenagas95-Dy-structure}
that of \textit{R }= Tm is incommensurate modulated structure with
$\overrightarrow{q}\mathrm{\simeq}$%
(0.093,0.093,0),\cite{Lynn97-RNi2B2C-ND-mag-crys-structure,Chang96-TmNi2B2C-mag-struct,Sternlieb97-TmNi2B2C}
and those of \textit{R }= Er, Tb, Gd are incommensurate modulated states with
$\overrightarrow{q}\mathrm{\simeq(}$0.55,0,0)$.$%
\cite{Lynn97-RNi2B2C-ND-mag-crys-structure,Sinha95-ErNi2b2C-mag-stru,Zarestky95-ErNi2B2C-mag-structure,Dervenagas96-mag-struct-TbNi2B2C,Detlefs99-Er-orthorhombic}
At $T_{\mathrm{WF}}$ (which is
%TCIMACRO{\TEXTsymbol{<}}%
%BeginExpansion
$<$%
%EndExpansion
$T_{\mathrm{N}}$), each of the latter states transforms into an
equal-amplitude, squared-up state and in the particular cases of \textit{R }=
Er and Tb, this transformation leads to a surge of a weak ferromagnetic
component.\cite{Canfield96-Er-WF-HT-diagram,Detlefs03-TbNi2B2C-ErNi2B2C-WF-Lockin,Kawano02-ErNi2B2C-neutron,Choi01-ErNi2B2C-neutron}
It is an\ experimental fact that none of the Ni-based compounds manifests a
ferromagnetic, FM, ground state though there is, at least for \textit{R }= Pr,
Ho, Dy, a strong FM intralayer coupling.

It is recalled that the Ni subsystem in \textrm{YNi}$_{\text{\textrm{2}}}%
$\textrm{B}$_{\text{\textrm{2}}}$\textrm{C} is magnetically
inactive\cite{Muller01-interplay-review} while the Co subsystem in
\textrm{YCo}$_{\text{\textrm{2}}}$\textrm{B}$_{\text{\textrm{2}}}$\textrm{C}
manifests an exchange enhanced paramagnetism:\cite{04-Pr(CoNi)2B2C} based on
the above model, these features indicate that $E_{F}$ in the Ni-based compound
is above a filled 3$d$ band and, in addition, is not at a steep region of
$N$($E$) while for the Co-based compound the curvature of $N$($E$) at $E_{F}$
must be positive and nonnegligible. Indeed electronic structure
calculations\cite{Coehoorn94-RNi2B2C-electronic-structure,Lee94-electronic-structure,Matthias94-electronic-structure,Pickett94-electronic-structure}
on $R$\textrm{Ni}$_{\text{\textrm{2}}}$\textrm{B}$_{\text{\textrm{2}}}%
$\textrm{C} ($R$ = Lu,Y) showed that $E_{F}$ is situated at the top of a
pronounced and narrow $N$($E$) peak and this peak lies on the top edge of
nearly filled Ni(3\textit{d}) bands: this, together with the intermediate
electron-phonon coupling and the smaller Stoner factor, explains the surge of
the superconductivity as well as the nonmagnetic character of the
Ni-sublattice. Furthermore, in case of a FM order, such an $N$($E$) peak at
$E_{F}$ would be exchange split due to the direct intra-atomic 4f-5d exchange
interaction:\cite{Coehoorn94-RNi2B2C-electronic-structure} such a split would
be higher than the superconducting gap; it is an experimental fact that, due
to their AFM structures, none of \ the magnetic $R$\textrm{Ni}%
$_{\text{\textrm{2}}}$\textrm{B}$_{\text{\textrm{2}}}$\textrm{C}
superconductors shows this splitting.

Band structure calculations\cite{Coehoorn94-RNi2B2C-electronic-structure} on
\textrm{LuCo}$_{\text{\textrm{2}}}$\textrm{B}$_{\text{\textrm{2}}}$\textrm{C}
showed that, for such isomorphous\ borocarbides, the rigid band model yields a
reasonable description of the band filling. Furthermore, $E_{F}$ is situated
at the decreasing but right-hand side of one of the peaks that receives a
considerable contribution from the Co 3$d$-band, and that $N$($E_{F}$) is of
the same magnitude as that of \textrm{LuNi}$_{\text{\textrm{2}}}$%
\textrm{B}$_{\text{\textrm{2}}}$\textrm{C}: the latter finding is consistent
with the observation that the Sommerfeld linear specific heat coefficients of
\textrm{YCo}$_{\text{\textrm{2}}}$\textrm{B}$_{\text{\textrm{2}}}$\textrm{C}
and \textrm{YNi}$_{\text{\textrm{2}}}$\textrm{B}$_{\text{\textrm{2}}}%
$\textrm{C} are equal.\cite{00-RCo2B2C} Furthermore, as that the first
derivatives of $N$($E$) are nonnegligible, then we expect the Co sublattice,
similar to $R$\textrm{Co}$_{2}$, to develop an intermediate character (or even
to be polarized) if $H_{\text{eff}}^{\text{Co}}>H_{\text{cr}}^{\text{Co}}$:
the intermediate character is indeed observed in \textrm{YCo}%
$_{\text{\textrm{2}}}$\textrm{B}$_{\text{\textrm{2}}}$\textrm{C}%
;\cite{01-PrDyCo2B2C,04-Pr(CoNi)2B2C} the possibility of Co-subsystem
polarization, on the other hand, would be addressed in this work. As far as
the $R$-sublattice magnetism is concerned, it is expected that the difference
in the lattice parameters and in the electronic band structure of the Co- and
Ni-based borocarbides would imply a modification in the character of the
mediating RKKY-type interactions and consequently in the character of their
magnetic ground states.

In this paper, we report on the extensive magnetic characterization of
\textrm{TbCo}$_{2}$\textrm{B}$_{2}$\textrm{C}. The successful synthesis of a
single-crystal sample of \textrm{TbCo}$_{2}$\textrm{B}$_{2}$\textrm{C} made it
possible to identify unambiguously the paramagnetic as well as the
ordered-state properties of the Tb sublattice: the former is dominated by a
Curie-Weiss behavior while the latter is found to be a FM structure with the
easy axis lying along the $a$ direction. Such a FM state is in sharp contrast
to the AFM-type mode of the isomorphous \textrm{TbNi}$_{2}$\textrm{B}$_{2}%
$\textrm{C.}%
\cite{Cho96-TbNi2B2C-anistropy-WF,Tommy96-TbNi2B2C-reorientation,Dervenagas96-mag-struct-TbNi2B2C,Kreyssig03-Mag-order-TbNi2B2C,Detlefs03-TbNi2B2C-ErNi2B2C-WF-Lockin,Song01-Tb-highResol-Mag-Xray}
This work also addressed the question\ of whether the surge of this FM state
(with a strong Tb moment and consequently a strong $H_{\text{eff}}^{\text{Co}%
}$) is able to polarize the Co 3\textit{d} subsystem. Our results suggest that
in spite of the intermediate character of the Co 3$d$ subsystem, the induced
$H_{\text{eff}}^{\text{Co}}$ is not able to bring about an unambiguous
spontaneous polarization.

\section{Experiment}

99.5\% $^{11}$B enriched polycrystals of \textrm{TbCo}$_{\mathrm{2}}%
$\textrm{B}$_{\mathrm{2}}$\textrm{C} were prepared by conventional arc-melt
method. These polycrystals were used as feeding rods during a floating-zone
synthesis, a process that we used for single crystals
growth.\cite{Takeya01-ErNi2B2C-substitution}

Characterization were carried out using magnetization [$M(T,H)$, extraction
method within the ranges 1.9 $\leq T\leq$ 300 K and $H\leq$ 90 kOe] and
zero-field specific heat [$C(T)$, semi-adiabatic method within the range 0.5
$\leq T\leq$ 15 K, accuracy better than 4\%]. A high-accuracy capacitance
dilatometer\cite{07-TbNi2B2C} was used for measuring the thermal expansion or
forced magnetostriction with a resolution better\ than 1 \AA . The relative
change in length, measured along the cosine directions $(\beta_{1}\beta
_{2}\beta_{3})$ when a field is applied along the cosine direction $(xyz)$, is
denoted as $^{\beta_{1}\beta_{2}\beta_{3}}\lambda_{xyz}(T)=$ $\left[
l(T,H)-l_{0}(T_{0},H_{0}))\right]  /l_{0}(T_{0},H_{0})$.

Neutron-diffraction measurements were carried out at the Institut
Laue-Langevin in Grenoble, France. Measurements were carried out on a powdered
as-prepared arc-melt polycrystalline sample (since large quantity is
desirable, see \S \ III.D below) as well as on a single crystal sample. The
powder diffraction patterns were collected within a temperature range 2 to 40
K using the D1B diffractometer with a selected incident wavelength of 2.42 $%
%TCIMACRO{\unit{\mathring{A}}}%
%BeginExpansion
\operatorname{\mathring{A}}%
%EndExpansion
$; Rietveld refinements of both crystallographic and magnetic structures\ were
carried out using the \textsc{Fullprof} package of Rodriguez-Carvajal
(www.ill.fr/dif/Soft/fp). Single crystal diffraction studies were performed on
the D10 four-circle diffractometer with $\lambda=2.3606$ $%
%TCIMACRO{\unit{\mathring{A}}}%
%BeginExpansion
\operatorname{\mathring{A}}%
%EndExpansion
$ over a wide range of $q$ space and within the temperature range 1.7 $<T<8$K.

\section{Results}

\subsection{Magnetization}%

%TCIMACRO{\FRAME{fthFU}{3.7178in}{4.811in}{0pt}{\Qcb{(Color online)
%$T$-dependent $\chi_{\text{dc}}$ (left-hand ordinate) and $\chi_{\text{dc}%
%}^{\text{-1}}$ (right-hand ordinate) curves of \QTR{rm}{TbCo}$_{2}%
%$\QTR{rm}{B}$_{2}$\QTR{rm}{C} measured along the \QTR{it}{a} and \QTR{it}{c}
%axes.\ The solid lines represent $\chi=C/(T-\theta)$; $C$ and $\theta$ are the
%Curie-Weiss constants. The large symbols on the $\chi_{\text{dc}}^{\text{-1}%
%}(T)$ curves represent the H/\QTR{it}{M(}$T,H\rightarrow0)$ as obtained from
%the Arrot plot. The upper-left inset shows, on expanded scales, the
%low-temperature $\chi_{\text{dc}}(T,H//a)$ for $H=$1 kOe and 10 kOe. The
%lower-right inset shows the magnetization isotherms at $T$=1.9 K which
%demonstrates that for $H\geq$2 kOe, there is no hysteresis effect (see
%text).}}{\Qlb{FM-Fig.1}}{FM-Fig1.eps}{\special{ language "Scientific Word";
%type "GRAPHIC";  maintain-aspect-ratio TRUE;  display "ICON";
%valid_file "F";  width 3.7178in;  height 4.811in;  depth 0pt;
%original-width 7.6614in;  original-height 9.9324in;  cropleft "0";
%filename 'TbCo2B2C-FM-Fig1.eps';file-properties "XNPEU";}}}%
%BeginExpansion
\begin{figure}
[th]
\begin{center}
\includegraphics[
height=4.811in,
width=3.7178in
]%
{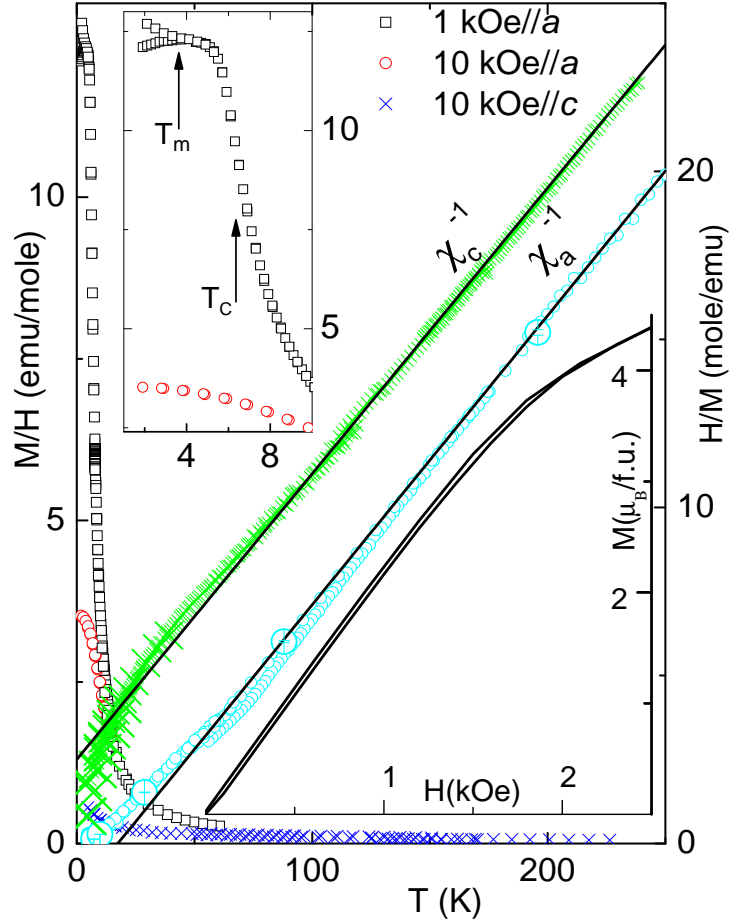}%
\caption{(Color online) $T$-dependent $\chi_{\text{dc}}$ (left-hand ordinate)
and $\chi_{\text{dc}}^{\text{-1}}$ (right-hand ordinate) curves of
\textrm{TbCo}$_{2}$\textrm{B}$_{2}$\textrm{C} measured along the \textit{a}
and \textit{c} axes.\ The solid lines represent $\chi=C/(T-\theta)$; $C$ and
$\theta$ are the Curie-Weiss constants. The large symbols on the
$\chi_{\text{dc}}^{\text{-1}}(T)$ curves represent the H/\textit{M(}%
$T,H\rightarrow0)$ as obtained from the Arrot plot. The upper-left inset
shows, on expanded scales, the low-temperature $\chi_{\text{dc}}(T,H//a)$ for
$H=$1 kOe and 10 kOe. The lower-right inset shows the magnetization isotherms
at $T$=1.9 K which demonstrates that for $H\geq$2 kOe, there is no hysteresis
effect (see text).}%
\label{FM-Fig.1}%
\end{center}
\end{figure}
%EndExpansion
Figure \ref{FM-Fig.1} shows the $T$-dependent $\chi_{\text{dc}}$ and
$\chi_{\text{dc}}^{\text{-1}}$ curves that were measured at different fields
and along the two principal tetragonal axes. The large symbols on the
$\chi_{\text{dc}}^{\text{-1}}(T)$ curve represent the \textit{H/M(}%
$T,H\rightarrow0)$ obtained from the Arrot plot: the excellent agreement
emphasizes that, within this temperature range, the contribution of magnetic
impurities is negligible. A Curie-Weiss fit down to liquid-nitrogen
temperatures of $\chi_{\text{dc}}(T,H\Vert c)$ gives $\mu_{\text{eff}}=$
9.7(1)$\mu_{\text{B}}$ and $\theta_{c}$= -29.4(1) K while that of
$\chi_{\text{dc}}(T,H\Vert a)$ gives $\mu_{\text{eff}}=$ 9.6(1)$\mu_{\text{B}%
}$ and $\theta_{a}$=14.4(1) K. Evidently, there are anisotropic forces but the
effective moments are in excellent agreement with the value expected for a
free Tb$^{3+}$ ion. Based on these anisotropic $\theta$ values, the first
Stevens coefficient in the crystal field description of a tetragonal symmetry
is estimated to be $B_{2}^{0}=$ 0.88(2) K; this compares well in sign and
magnitude with that of \textrm{TbNi}$_{2}$\textrm{B}$_{2}$\textrm{C}
($B_{2}^{0}$ =1.2(1) K).\cite{07-TbNi2B2C} This similarity suggests that the
crystalline electric field, CEF, at the Tb$^{3+}$ site of both isomorphs are
similar: indeed both sites have the same $D_{4h}$ symmetry and almost the same
charge distribution. On lowering the temperatures toward liquid helium region,
$\chi_{\text{dc}}(T,H\Vert a)$\ increases relatively fast and afterward tends
toward saturation. Considering the characteristic magnetic features manifested
in the magnetization, specific heat, and neutron diffraction (see below), this
fast increase (which is followed by saturation) is caused by the process of
approaching and the eventual onset of a FM order wherein the moments point
along the \textit{a }axis: $T_{\text{C}}$= 6.3(2) K is the point of maximum inclination.

The upper-left inset of Fig. \ref{FM-Fig.1} shows, on an expanded scale, a
magnetic hysteresis occurring at $T$ $\leq T_{\text{m}}=$3.7(2) K and $H$=1
kOe. The $M(H\Vert a,$ 1.9 K$)$ curve, shown in the lower-right inset of Fig.
\ref{FM-Fig.1}, reveals that this hysteresis effect disappears for $H\geq$ 2
kOe. Since this $T_{\text{m}}$-feature is sample-dependent (see below), it is
attributed to a contaminating magnetic phase, the magnetization of which
saturates completely to 0.05\ $\mu_{\text{B}}/$formula unit for field higher
than 2 kOe. Based on the weight ratios of the magnetic moments, the fraction
of the Tb ions in this spurious phase relative to the major \textrm{TbCo}%
$_{2}$\textrm{B}$_{2}$\textrm{C} phase is estimated to be 0.7\%.%

%TCIMACRO{\FRAME{fthFU}{3.5578in}{4.8127in}{0pt}{\Qcb{(Color online)
%Magnetization isotherms of \QTR{rm}{TbCo}$_{2}$\QTR{rm}{B}$_{2}$\QTR{rm}{C} at
%different temperatures and for different field orientations. Two different
%set-ups were employed one with a field up to 90 kOe and another up to 50 kOe.
%The high-field magnetization evolves as $M(H\Vert a,$ 2K$)=\mu_{\text{sp}%
%}+\chi_{\text{hf}}H$ and is represented by the solid line (see text).}}%
%{\Qlb{FM-Fig.2}}{FM-Fig.2}{\special{ language "Scientific Word";
%type "GRAPHIC";  maintain-aspect-ratio TRUE;  display "ICON";
%valid_file "F";  width 3.5578in;  height 4.8127in;  depth 0pt;
%original-width 7.3414in;  original-height 9.9471in;  cropleft "0";
%filename 'TbCo2B2C-FM-Fig2.eps';file-properties "XNPEU";}}}%
%BeginExpansion
\begin{figure}
[th]
\begin{center}
\includegraphics[
height=4.8127in,
width=3.5578in
]%
{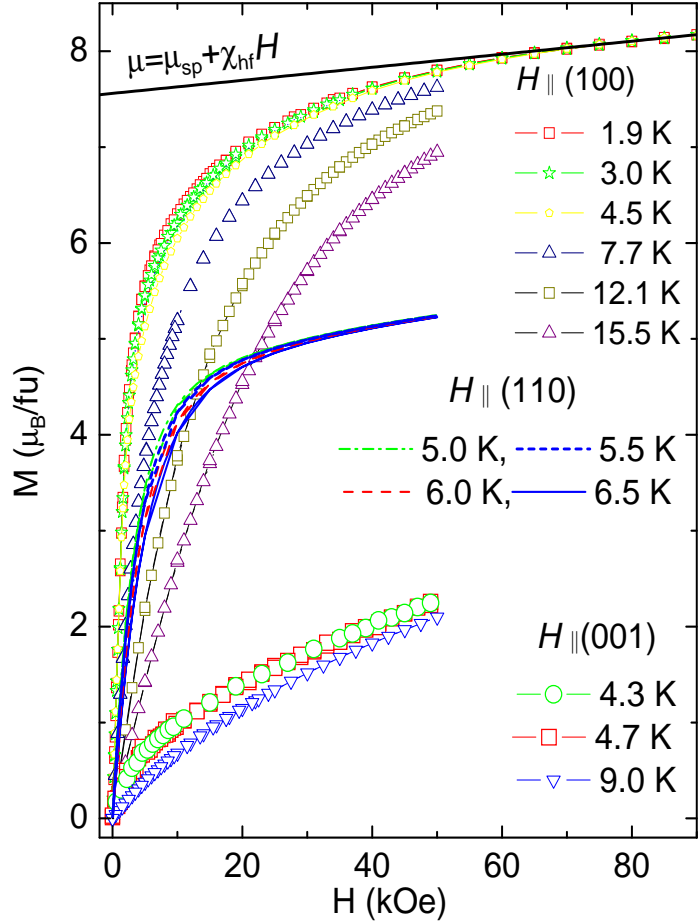}%
\caption{(Color online) Magnetization isotherms of \textrm{TbCo}$_{2}%
$\textrm{B}$_{2}$\textrm{C} at different temperatures and for different field
orientations. Two different set-ups were employed one with a field up to 90
kOe and another up to 50 kOe. The high-field magnetization evolves as
$M(H\Vert a,$ 2K$)=\mu_{\text{sp}}+\chi_{\text{hf}}H$ and is represented by
the solid line (see text).}%
\label{FM-Fig.2}%
\end{center}
\end{figure}
%EndExpansion

Figure \ref{FM-Fig.2} confirms the above-mentioned magnetic anisotropy, due to
which the \textit{a }(\textit{c}) is the easy (hard) axis. Within the
studied\ ranges of $H$ and $T$, the magnetization isotherms do not show any
field-induced transition, rather, only a monotonic and steady increase
(tending towards saturation) which is characteristic of a forced domain
alignment: this supports the earlier inference of a FM order. Furthermore, the
high-field magnetization increases as $M(H\Vert a,$ 2K$)=\mu_{\text{sp}}%
+\chi_{\text{hf}}H\ $[$\mu_{\text{sp}}$=7.6(3)$\mu_{\text{B}}$, $\chi
_{\text{hf}}$=6.8(4)$\times$10$^{-6}$ $\mu_{\text{B}}$/Oe corresponding to
38(2)$\times$10$^{-3}$ emu/mol] attaining $\mu$(90 kOe, 2 K) = 8.2$\mu
_{\text{B}}$. $\mu_{\text{sp}}$ is only 3\% lower than the reported moment of
\textrm{TbNi}$_{2}$\textrm{B}$_{2}$\textrm{C} ( $\mu_{\text{Tb}}=$ 7.78
$\mu_{\text{B}}$%
,\cite{Lynn97-RNi2B2C-ND-mag-crys-structure,Cho96-TbNi2B2C-anistropy-WF}) but
16\% lower than the one expected for a free Tb$^{3+}$ ion: this observed
moment lowering confirms the above-mentioned influence of the CEF effects and
that these effects are similar to the ones observed in the Ni-based isomorph.
Alternatively, let us assume that the difference between $\mu_{\text{sp}}$ of
\textrm{TbCo}$_{2}$\textrm{B}$_{2}$\textrm{C} and the reported moment of
\textrm{TbNi}$_{2}$\textrm{B}$_{2}$\textrm{C} is due exclusively to the
spontaneous polarization of the Co 3\textit{d} orbitals which are coupled
ferrimagnetically to the FM Tb sublattice. Then, based on the relation
$\mu_{\text{sp}}(\mathrm{TbCo}_{2}\mathrm{B}_{2}\mathrm{C)}=$ $\mu_{\text{sp}%
}(\mathrm{TbNi}_{2}\mathrm{B}_{2}\mathrm{C)}$ $-2.\mu_{\text{Co}}$, the
maximum possible Co moment would be $\mu_{\text{Co}}=$0.25 $\mu_{\text{B}}$.
Such a $\mu_{\text{Co}}$ value is surprisingly small, amounting to only 0.25
hole/Co atom in the 3\textit{d} band; this value is almost one fourth of the
Co moment encountered in the heavy members of the $R$\textrm{Co}$_{2}$
series,\cite{Cyrot79-electr-cal} but it is similar to the value observed in
the low spin state of Co in, e.g., \textrm{ErCo}$_{2}$%
.\cite{Liu06-Co-moment-ErCo2,Herrero-Albillos07-ErCo2-paramagnetism} It is
more likely that $H_{\text{eff}}^{\text{Co}}$ $<$ $H_{\text{cr}}^{\text{Co}}$
(see \S \ IV) since if there is any spontaneous Co polarization then the
induced moment should be much higher than 0.25 $\mu_{\text{B}}$. As\ the\ CW
law describes well the paramagnetic susceptibility of $\mathrm{TbCo}%
_{2}\mathrm{B}_{2}\mathrm{C}$ (see above), then the exchange-enhancement
factor for the Co-subsystem susceptibility must be extremely small.

\subsection{Specific heat}%

%TCIMACRO{\FRAME{fthFU}{3.6106in}{4.8127in}{0pt}{\Qcb{(Color online) Zero-field
%magnetic specific heat (symbol) and calculated magnetic entropy (dashed line)
%of single-crystal \QTR{rm}{TbCo}$_{2}$\QTR{rm}{B}$_{2}$\QTR{rm}{C}. The
%nuclear, electronic, and lattice contributions were already subtracted (the
%latter two were obtained from \QTR{rm}{YCo}$_{2}$\QTR{rm}{B}$_{2}%
%$\QTR{rm}{C).}\cite{00-RCo2B2C,04-Pr(CoNi)2B2C}\ The solid line represents a
%fit to the expression of the magnon contribution (Eq. \ref{Cm-FM}) assuming a
%FM order of the Tb sublattice. The inset (with the same scales as the main
%panel) compares the single-crystal magnetic specific heat (solid line) with
%that of a polycrystalline sample (symbol); the observed difference reflects a
%strong dependence on the sample form, contamination, and history.
%Nevertheless, their magnetic entropies approach each other for $T>T_{\text{c}%
%}$ confirming, as it should, the conservation of the total entropy (see
%text).}}{\Qlb{FM-Fig.3}}{FM-Fig.3}{\special{ language "Scientific Word";
%type "GRAPHIC";  maintain-aspect-ratio TRUE;  display "ICON";
%valid_file "F";  width 3.6106in;  height 4.8127in;  depth 0pt;
%original-width 7.4382in;  original-height 9.935in;  cropleft "0";
%filename 'TbCo2B2C-FM-Fig3.eps';file-properties "XNPEU";}}}%
%BeginExpansion
\begin{figure}
[th]
\begin{center}
\includegraphics[
height=4.8127in,
width=3.6106in
]%
{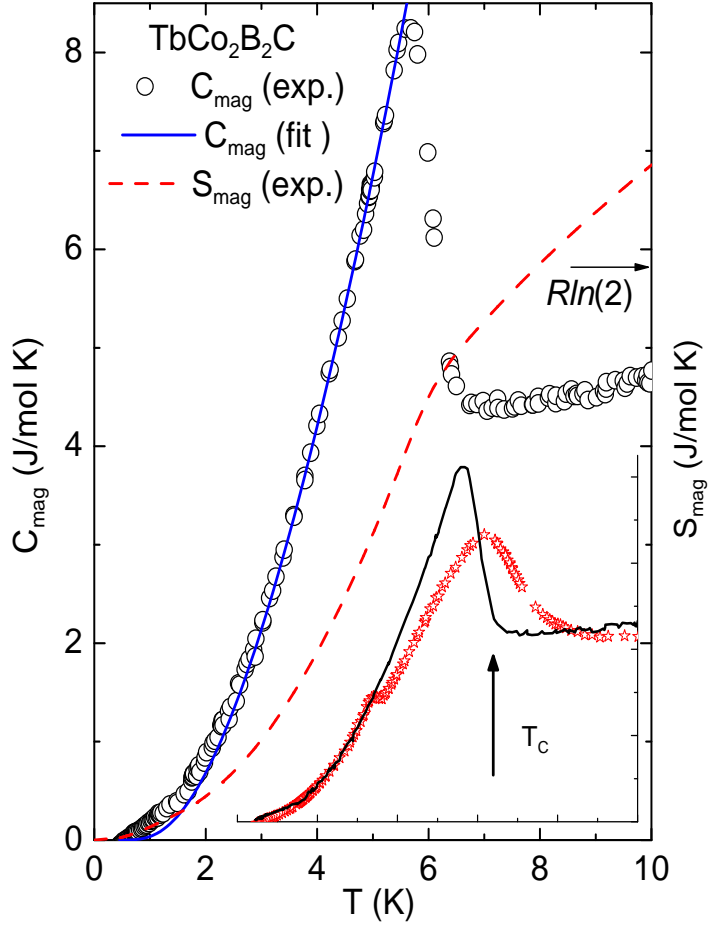}%
\caption{(Color online) Zero-field magnetic specific heat (symbol) and
calculated magnetic entropy (dashed line) of single-crystal \textrm{TbCo}%
$_{2}$\textrm{B}$_{2}$\textrm{C}. The nuclear, electronic, and lattice
contributions were already subtracted (the latter two were obtained from
\textrm{YCo}$_{2}$\textrm{B}$_{2}$\textrm{C).}%
\cite{00-RCo2B2C,04-Pr(CoNi)2B2C}\ The solid line represents a fit to the
expression of the magnon contribution (Eq. \ref{Cm-FM}) assuming a FM order of
the Tb sublattice. The inset (with the same scales as the main panel) compares
the single-crystal magnetic specific heat (solid line) with that of a
polycrystalline sample (symbol); the observed difference reflects a strong
dependence on the sample form, contamination, and history. Nevertheless, their
magnetic entropies approach each other for $T>T_{\text{c}}$ confirming, as it
should, the conservation of the total entropy (see text).}%
\label{FM-Fig.3}%
\end{center}
\end{figure}
%EndExpansion

Figure \ref{FM-Fig.3} shows the zero-field magnetic specific heat and entropy
of single-crystal \textrm{TbCo}$_{2}$\textrm{B}$_{2}$\textrm{C} obtained after
subtracting the nuclear, electronic, and lattice contributions (the latter two
were obtained from \textrm{YCo}$_{2}$\textrm{B}$_{2}$\textrm{C).}%
\cite{00-RCo2B2C,04-Pr(CoNi)2B2C}\ The nuclear contribution is of dominant
importance only at very low temperatures and was evaluated from the
diagonalization of the hyperfine Hamiltonian.\cite{Kruis-Nclr-Ho-Tb-Schottky}
It is worth mentioning that at very low-temperature, the nuclear contribution
is much stronger than the magnetic one: as such the propagation of errors due
to successive subtraction of non-magnon contributions would eventually
influence the absolute value of $C_{M}(T)$; this may undermine the quality of
the comparison between the theoretical and experimental magnon contributions
(see Fig. \ref{FM-Fig.3}).

Both $C_{mag}$($T$) and $S_{mag}$($T$) curves do confirm the onset of the
magnetic order at\ $T_{c}$: the former curve rises very sharply at $T_{c}$
while the latter manifests a pronounced change of slope. Considering the
magnetic structure to be associated with the FM\ order of the Tb-sublattice
(see below), we fit the experimental $C_{mag}$($T$) to the theoretical magnon
expression of\ Eq. \ref{Cm-FM} (see Appendix): as can be seen in Fig.
\ref{FM-Fig.3}, the excellent fit to Eq. \ref{Cm-FM} gives the spin-wave
stiffness coefficient $D$=26.3(5) K and the gap parameter $\Delta$=8.4(2) K.
The high value of $D$ is indicative of stronger effective exchange couplings.
On the other hand, the value of $\Delta$ (which from Eq. \ref{Delta} is a
measure of the anisotropic field) is consistent with the strong anisotropic
features observed in the magnetization measurements.

The inset of Fig. \ref{FM-Fig.3} compares the\ measured $C_{mag}$($T$) of a
single-crystal sample with that of a polycrystalline one. Evidently the
$T_{\text{m}}$-transition is sample-dependent: while $C_{mag}$($T$) of the
polycrystalline sample manifests a pronounced event at $T_{\text{m}}$=3.6(2)
K, that of the single-crystal hardly shows any anomaly. As mentioned above,
the single-crystal sample contains only a 0.7\% spurious phase and this
concentration limit is lower than the resolution of the specific heat set-up.

\subsection{magnetostriction}

\begin{figure}
[th]
\begin{center}
\includegraphics[
height=4.8127in,
width=3.3512in
]%
{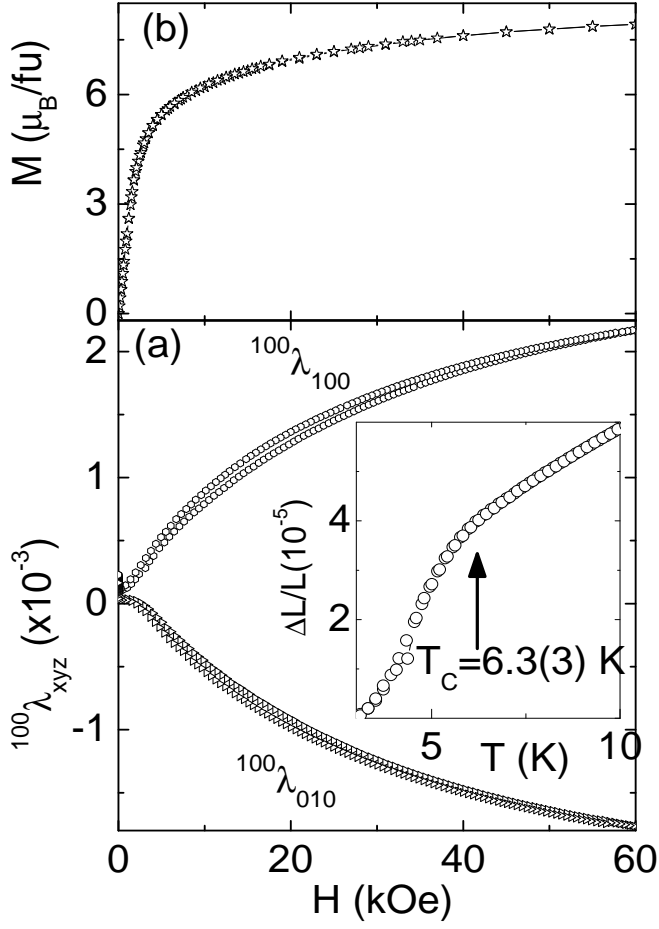}%
\caption{Representative forced magnetostrictions (a) $^{100}\lambda_{100}(H$,
3K$)$ and $^{100}\lambda_{010}(H$, 3K$)$ curves of \textrm{TbCo}$_{2}%
$\textrm{B}$_{2}$\textrm{C} are compared to (b) the isothermal magnetization
at 3 K. The magnetostriction curves are given relative to their zero-field
values. The inset shows the thermal evolution of the zero-field $\Delta L/L$
measured along the $a$ axis. The arrow marks the $T_{c}$ value which was
determined from the specific heat measurement of Fig.3. }%
\label{FM-Fig.4}%
\end{center}
\end{figure}

Figure \ref{FM-Fig.4} shows the forced magnetostriction isotherms measured
along the $a$ and $b$ axis with $H\mathrm{\Vert}a$. Once more (apart from the
low-field, domain-wall sweeping-out region and the saturated regime) there is
no metamagnetic transition in these isotherms. The inset of Fig.
\ref{FM-Fig.4} indicates clearly that below $T_{c}$, \textrm{TbCo}$_{2}%
$\textrm{B}$_{2}$\textrm{C}\ undergoes a spontaneous distortion which - based
on Fig. \ref{FM-Fig.4}(a), see also \S \ III.D - is attributed to an
orthorhombic distortion of the tetragonal unit cell. A similar distortion was
reported for \textrm{TbNi}$_{2}$\textrm{B}$_{2}$\textrm{C}%
.\cite{Song99-Tb-magnetostriction,Song01-Tb-highResol-Mag-Xray,Song01-Tb-dichroism-HRMXRD,Detlefs03-TbNi2B2C-ErNi2B2C-WF-Lockin,07-TbNi2B2C}
Then the behavior of the forced magnetostriction of Fig.\ref{FM-Fig.4} can be
understood as follows: at zero-field, there is an equal distribution of
domains along each of the $a$ and $b$ axes; an applied field along, say, the
$a$ axis would involve a rearrangement of the orthorhombically-distorted
domains and as such induces an increase in $M(H\Vert a)$ [Fig. \ref{FM-Fig.4}
(b)], an increase in $^{100}\lambda_{100}(H)$, and a decrease in
$^{010}\lambda_{100}(H)$ [Fig. \ref{FM-Fig.4} (a)].%

%TCIMACRO{\FRAME{fthFU}{3.3512in}{4.8127in}{0pt}{\Qcb{Representative forced
%magnetostrictions (a) $^{100}\lambda_{100}(H$, 3K$)$ and $^{100}\lambda
%_{010}(H$, 3K$)$ curves of \QTR{rm}{TbCo}$_{2}$\QTR{rm}{B}$_{2}$\QTR{rm}{C}
%are compared to (b) the isothermal magnetization at 3 K. The magnetostriction
%curves are given relative to their zero-field values. The inset shows the
%thermal evolution of the zero-field $\Delta L/L$ measured along the $a$ axis.
%The arrow marks the $T_{c}$ value which was determined from the specific heat
%measurement of Fig.3. }}{\Qlb{FM-Fig.4}}{FM-Fig.4}%
%maintain-aspect-ratio TRUE;  display "ICON";  valid_file "F";
%width 3.3512in;  height 4.8127in;  depth 0pt;  original-width 7.4512in;
%original-height 10.7263in;  cropleft "0";  croptop "1";  cropright "1";
%cropbottom "0";  filename 'TbCo2B2C-FM-Fig4.eps';file-properties "XNPEU";}}}%
%BeginExpansion
%EndExpansion

\subsection{Neutron Diffraction}%

%TCIMACRO{\FRAME{fthFU}{3.3512in}{4.811in}{0pt}{\Qcb{(Color online)
%Representative neutron powder diffractograms of\ as-prepared polycrystalline
%\QTR{rm}{TbCo}$_{\QTR{rm}{2}}$\QTR{rm}{B}$_{\QTR{rm}{2}}$\QTR{rm}{C}. To aid
%in visualizing the magnetic modes, this plot is limited to temperatures\ below
%10 K and to scattering angle lower than 50$^{\circ}$. The diffractogram at 6.4
%K (denoted by thick solid line) can be taken as a demarcation between the
%paramagnetic and FM phases. }}{\Qlb{FM-Fig.5}}{FM-Fig.5}%
%maintain-aspect-ratio TRUE;  display "ICON";  valid_file "F";
%width 3.3512in;  height 4.811in;  depth 0pt;  original-width 7.6864in;
%original-height 11.0627in;  cropleft "0";  croptop "1";  cropright "1";
%cropbottom "0";  filename 'TbCo2B2C-FM-Fig5.eps';file-properties "XNPEU";}}}%
%BeginExpansion
\begin{figure}
[th]
\begin{center}
\includegraphics[
height=4.811in,
width=3.3512in
]%
{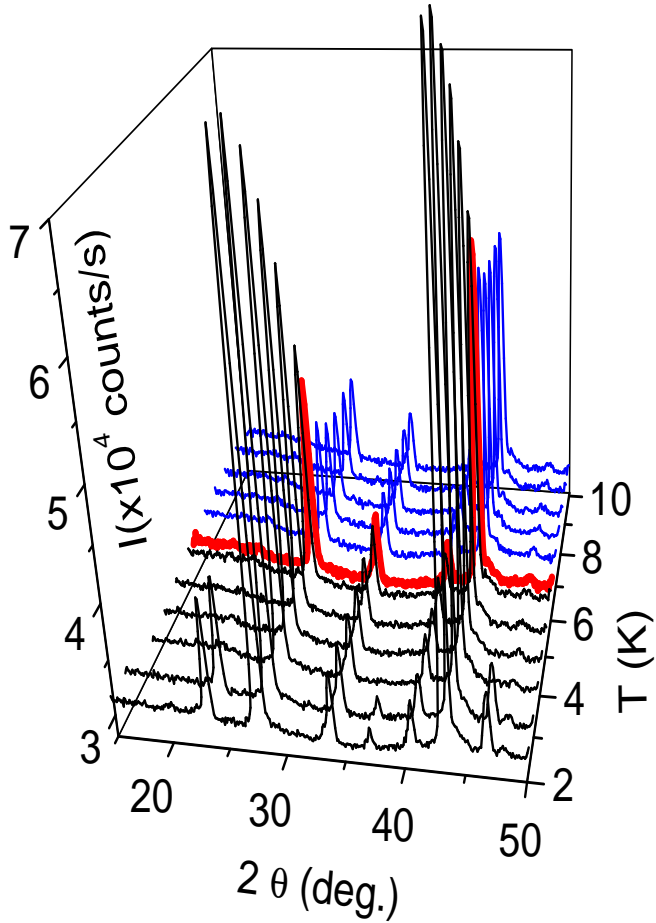}%
\caption{(Color online) Representative neutron powder diffractograms
of\ as-prepared polycrystalline \textrm{TbCo}$_{\mathrm{2}}$\textrm{B}%
$_{\mathrm{2}}$\textrm{C}. To aid in visualizing the magnetic modes, this plot
is limited to temperatures\ below 10 K and to scattering angle lower than
50$^{\circ}$. The diffractogram at 6.4 K (denoted by thick solid line) can be
taken as a demarcation between the paramagnetic and FM phases. }%
\label{FM-Fig.5}%
\end{center}
\end{figure}
%EndExpansion%
%TCIMACRO{\FRAME{fthFU}{3.2612in}{4.8127in}{0pt}{\Qcb{(Color online) The
%magnetic diffractograms of as-prepared polycrystalline
%\QTR{rm}{TbCo}$_{\QTR{rm}{2}}$\QTR{rm}{B}$_{\QTR{rm}{2}}$\QTR{rm}{C} sample.
%These were obtained after subtracting either the pattern at 10 K ($a$ and $b$)
%or the pattern at 4.2 K ($c$). ($a$) The FM mode; ($b$) the superposition of
%the FM and the magnetic contamination contribution at $T$=2.0 K (the strongest
%peaks of the later are denoted by vertical arrows); (c) the thermal evolution
%($T<T_{m}$) of the magnetic patterns relative to the one at 4.2K.}}%
%{\Qlb{FM-Fig.6}}{FM-Fig.6}{\special{ language "Scientific Word";
%type "GRAPHIC";  maintain-aspect-ratio TRUE;  display "ICON";
%valid_file "F";  width 3.2612in;  height 4.8127in;  depth 0pt;
%original-width 7.1451in;  original-height 10.5715in;  cropleft "0";
%filename 'TbCo2B2C-FM-Fig6.eps';file-properties "XNPEU";}}}%
%BeginExpansion
\begin{figure}
[th]
\begin{center}
\includegraphics[
height=4.8127in,
width=3.2612in
]%
{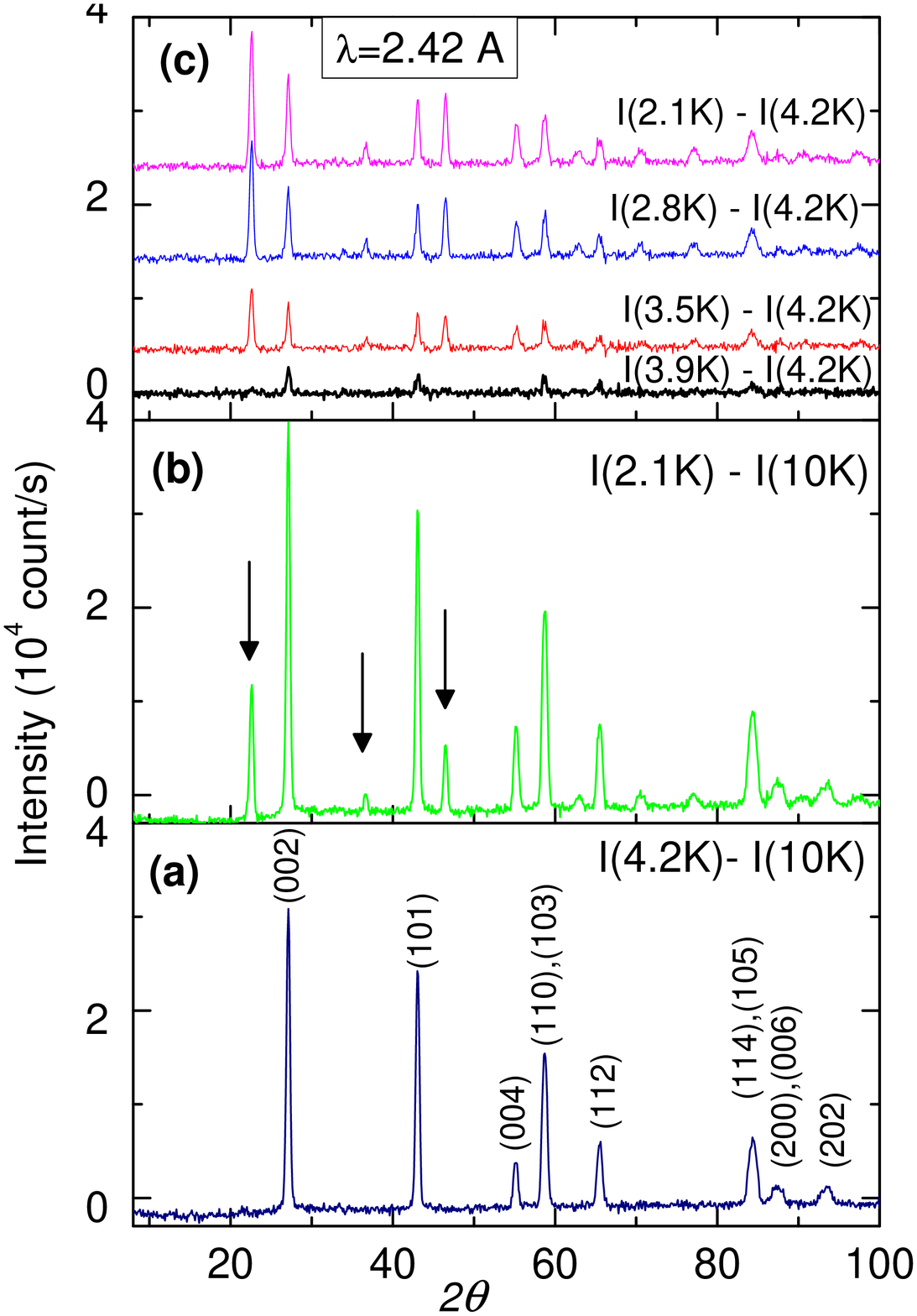}%
\caption{(Color online) The magnetic diffractograms of as-prepared
polycrystalline \textrm{TbCo}$_{\mathrm{2}}$\textrm{B}$_{\mathrm{2}}%
$\textrm{C} sample. These were obtained after subtracting either the pattern
at 10 K ($a$ and $b$) or the pattern at 4.2 K ($c$). ($a$) The FM mode; ($b$)
the superposition of the FM and the magnetic contamination contribution at
$T$=2.0 K (the strongest peaks of the later are denoted by vertical arrows);
(c) the thermal evolution ($T<T_{m}$) of the magnetic patterns relative to the
one at 4.2K.}%
\label{FM-Fig.6}%
\end{center}
\end{figure}
%EndExpansion

The thermal evolution of the powder diffractograms are shown Figs.
\ref{FM-Fig.5}-\ref{FM-Fig.6}. For $T>T_{\text{c}}$, the patterns consist of
the tetragonal crystal structure of \textrm{TbCo}$_{2}$\textrm{B}$_{2}%
$\textrm{C} and a small impurity phase. On the other hand, for $T_{\text{m}%
}<T<$ $T_{\text{c}}$, it is evident that the magnetic reflections are piled up
on the top of the nuclear Bragg peaks: a $q_{\text{0}}$=(000) mode.
Considering that the paramagnetic state is dominated by the Tb moment, that
the evolution of the isothermal magnetization and magnetostriction indicates
no metamagnetic transition which can be related to the onset of Co moment,
then this $q_{\text{0}}$ mode must be related to the Tb sublattice.
Furthermore, since the Tb ion occupies the special 2$a$ site in the unit cell,
then this mode must be FM: confirming the conclusions drawn from the
magnetization, magnetostriction, and specific heat studies. Indeed, Fig.
\ref{FM-Fig.6} (a) confirms that this pattern is a\ FM mode. Alternatively, if
this mode is related to the Co sublattice then, due to the multiplicity of the
4$d$ site occupied by the Co atoms, the magnetic order should be either AFM
(if only due to Co subsystem) or ferrimagnetic (if both subsystems are
ordered): in the light of all the above-mentioned results, both possibilities
must be ruled out (see also \S \ IV).

\begin{table}[th]
\caption{Comparison of the space groups, atomic positions, and isotropic
thermal parameters of \ TbCo$_{2}$B$_{2}$C and TbNi$_{2}$B$_{2}$C. The data on
TbNi$_{2}$B$_{2}$C are taken from Lynn \textit{et al}%
.\cite{Lynn97-RNi2B2C-ND-mag-crys-structure} The same thermal parameters
reported for TbNi$_{2}$B$_{2}$C are also used for the analysis of TbCo$_{2}%
$B$_{2}$C.}%
\label{Table I}%
$%
\begin{tabular}
[c]{ccccccc}\hline\hline
\multicolumn{3}{c}{} & Tb & Co & B & C\\\hline
TbNi$_{2}$B$_{2}$C & $P4/mmm$ & position & (000) & ($\frac{1}{2}0\frac{1}{4}%
$) & (0,0,$0.357$) & ($\frac{1}{2}\frac{1}{2}0$)\\
&  & thermal factor & 0.47 & 0.57 & 0.77 & 0.85\\
TbCo$_{2}$B$_{2}$C & $Immm$ & position & (000) & ($\frac{1}{2}0\frac{1}{4}$) &
(0,0,$0.354$) & ($\frac{1}{2}\frac{1}{2}0$)\\
&  & thermal factor & $\ 0.47$ & 0.57 & $\ 0.77$ & 0.85$\ $\\\hline\hline
\end{tabular}
\ \ \ \ \medskip$\end{table}

\begin{figure}
[th]
\begin{center}
\includegraphics[
height=4.8646in,
width=3.2846in
]%
{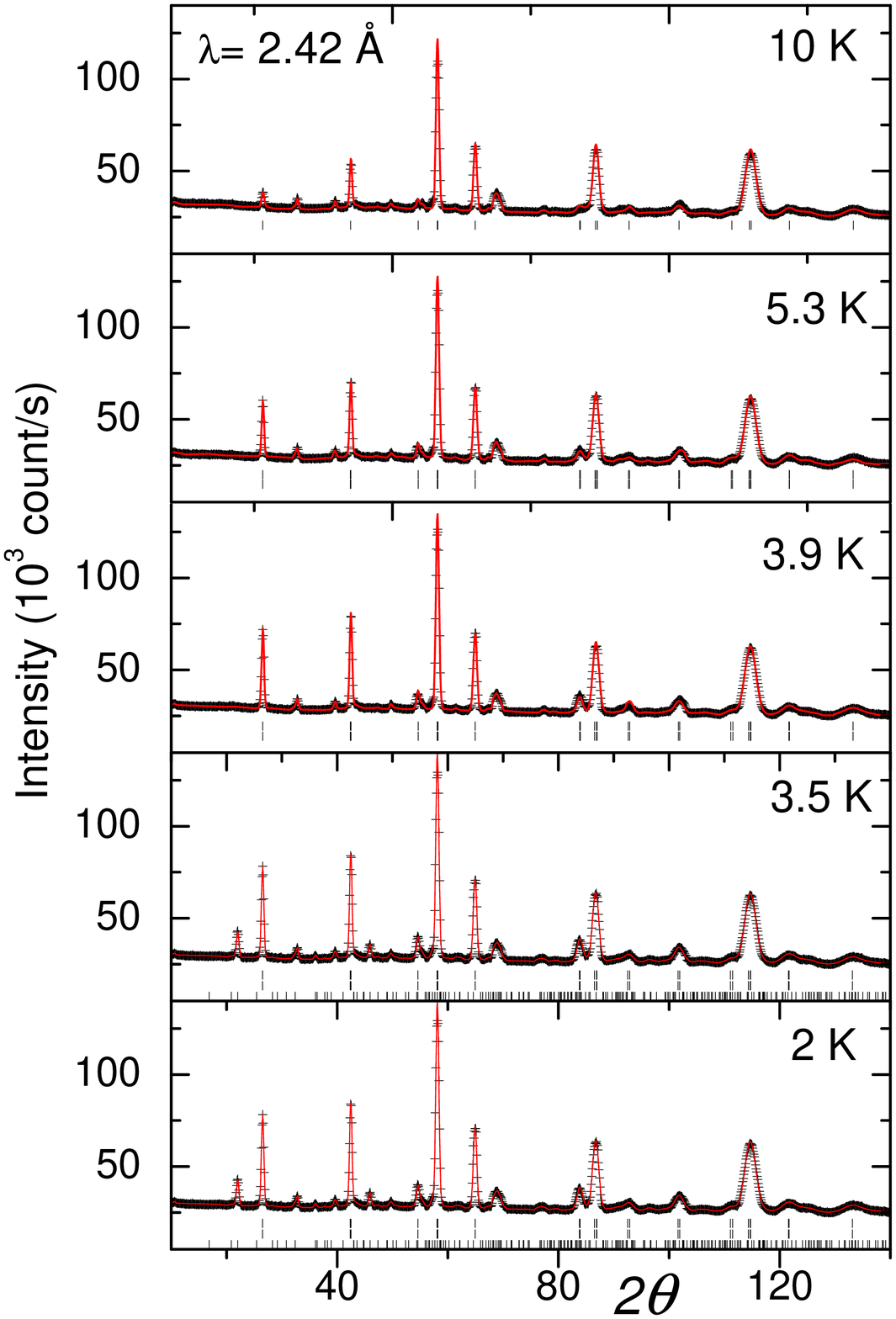}%
\caption{(Color online) Rietveld analysis on representative powder
diffractograms. Within the range $T_{\text{m}}<T<$ $T_{\text{c}},$ the total
contribution is composed of a nuclear and a FM component while for $T<$
$T_{\text{m}}$ it is a sum of three phases: a nuclear, a FM, and an impurity
phase (see text). The thermal evolution of the lattice parameters and magnetic
moment are given in Fig. \ref{FM-Fig.8}. The Bragg $R$-factor for the
structural phase clusters around 2 while for the magnetic phase around 2.1.}%
\label{FM-Fig.7}%
\end{center}
\end{figure}

\begin{figure}
[th]
\begin{center}
\includegraphics[
height=5.047in,
width=3.5405in
]%
{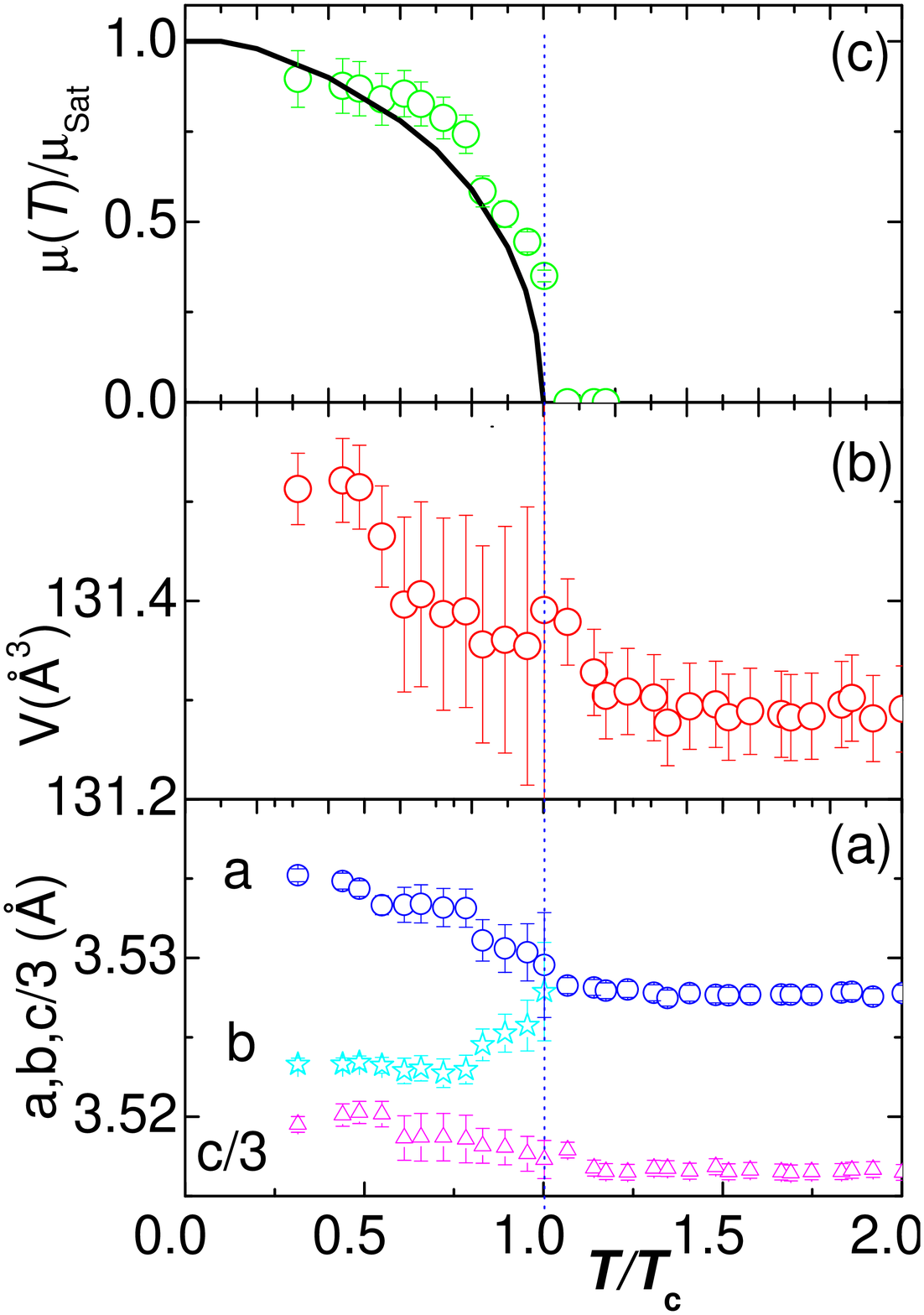}%
\caption{(Color on line) The lattice parameters (a - b) and the reduced
magnetic moment $\mu(T)/\mu(2$ K) versus the reduced temperature
\textit{T/T}$_{c}$ (c). These parameters were obtained from the Rietveld
analysis (see Fig. \ref{FM-Fig.7}). \ The solid line in the upper panel
represent the Brillouin function for \textit{J}=6.}%
\label{FM-Fig.8}%
\end{center}
\end{figure}

To extract more information, Rietveld analysis was carried out on the
diffractograms measured within the range $T_{\text{m}}<T<$ $T_{\text{c}}$.
Because of the structural distortion, we used the \textit{Immm} space group
together with the parameters given in Table \ref{Table I}. Representative
analyzed diffractograms are shown in Fig. \ref{FM-Fig.7} while the obtained
cell parameters and the magnetic moment are given in Fig. \ref{FM-Fig.8}. The
analysis indicates that the lattice parameters [Fig. \ref{FM-Fig.8} (a-b)]
undergoes a noticeable orthorhombic distortion below $T_{\text{c}}$ which is
consistent with the magnetoelastic effects observed in Fig. \ref{FM-Fig.4}.
Furthermore, the analysis revealed that the zero-field Tb magnetic moment is
along the longest side of the\ base of the orthorhombic cell.%

%TCIMACRO{\FRAME{fthFU}{3.2846in}{4.8646in}{0pt}{\Qcb{(Color online) Rietveld
%analysis on representative powder diffractograms. Within the range
%$T_{\text{m}}<T<$ $T_{\text{c}},$ the total contribution is composed of a
%nuclear and a FM component while for $T<$ $T_{\text{m}}$ it is a sum of three
%phases: a nuclear, a FM, and an impurity phase (see text). The thermal
%evolution of the lattice parameters and magnetic moment are given in Fig.
%\ref{FM-Fig.8}. The Bragg $R$-factor for the structural phase clusters around
%2 while for the magnetic phase around 2.1.}}{\Qlb{FM-Fig.7}}{FM-Fig.7}%
%maintain-aspect-ratio TRUE;  display "ICON";  valid_file "F";
%width 3.2846in;  height 4.8646in;  depth 0pt;  original-width 8.035in;
%original-height 10.4331in;  cropleft "0";  croptop "1";  cropright "1";
%cropbottom "0";  filename 'TbCo2B2C-FM-Fig7.eps';file-properties "XNPEU";}}}%
%BeginExpansion
%EndExpansion
%

%TCIMACRO{\FRAME{fthFU}{3.5405in}{5.047in}{0pt}{\Qcb{(Color on line) The
%lattice parameters (a - b) and the reduced magnetic moment $\mu(T)/\mu(2$ K)
%versus the reduced temperature \QTR{it}{T/T}$_{c}$ (c). These parameters were
%obtained from the Rietveld analysis (see Fig. \ref{FM-Fig.7}). \ The solid
%line in the upper panel represent the Brillouin function for \QTR{it}{J}=6.}}%
%{\Qlb{FM-Fig.8}}{FM-Fig.8}{\special{ language "Scientific Word";
%type "GRAPHIC";  maintain-aspect-ratio TRUE;  display "ICON";
%valid_file "F";  width 3.5405in;  height 5.047in;  depth 0pt;
%original-width 8.035in;  original-height 10.4884in;  cropleft "0";
%filename 'TbCo2B2C-FM-Fig8.eps';file-properties "XNPEU";}}}%
%BeginExpansion
%EndExpansion

Below $T_{\text{m}}\approx$3.7 K, the thermal evolution of the diffractograms
reveals two features [Figs. \ref{FM-Fig.6} (b and c)]: First, the intensity of
the FM mode evolves smoothly and independently. Secondly, there is a surge of
additional magnetic peaks [marked by the vertical arrows in the difference
plots of Fig. \ref{FM-Fig.6} (b and c)]. Since Figs. \ref{FM-Fig.1} to
\ref{FM-Fig.4} do not indicate any event that can be related to an
order-to-order transition of Tb magnetic order, then these peaks can not be
associated with the magnetic pattern of the Tb-sublattice. As these
diffractograms were collected on as-prepared, arc-melt polycrystalline sample
- which as mentioned above contains magnetic contamination - then these
additional peaks are taken to be due to the same magnetic contamination, which
is responsible for the hysteresis event in Fig. \ref{FM-Fig.1} and the weak
specific heat anomaly in the inset of Fig. \ref{FM-Fig.3} (we argue, in
\S \ IV, that these events can not be due to an onset of a Co itinerant
moment).%
%TCIMACRO{\FRAME{fthFU}{3.3442in}{4.6855in}{0pt}{\Qcb{The upper panel shows, on
%a three dimensional plot, the thermal evolution of the (0,0,4) peak and an
%impurity peak. The latter together with its Gaussian fit is shown, on an
%expanded scale, in the inset. The lower panel shows the powder diffractogram
%at 2 K (taken from Fig.7). The \ intensity of the (0,0,4) and impurity peaks
%(marked by a vertical arrow) are almost equal in the powder diffractrogram but
%are a factor of 20 different in the single-crystal diffractograms (no
%correction for multiplicity factors are considered). The wave length used for
%the measurement in the upper panel is 2.3606 $\unit{\mathring{A}}$ while for
%the lower panel is 2.42 $\unit{\mathring{A}}.$}}{\Qlb{FM-Fig.9}}%
%{FM-Fig9}{\special{ language "Scientific Word";  type "GRAPHIC";
%maintain-aspect-ratio TRUE;  display "ICON";  valid_file "F";
%width 3.3442in;  height 4.6855in;  depth 0pt;  original-width 7.5887in;
%original-height 10.6562in;  cropleft "0";  croptop "1";  cropright "1";
%cropbottom "0";  filename 'TbCo2B2C-FM-Fig9.eps';file-properties "XNPEU";}}}%
%BeginExpansion
\begin{figure}
[th]
\begin{center}
\includegraphics[
height=4.6855in,
width=3.3442in
]%
{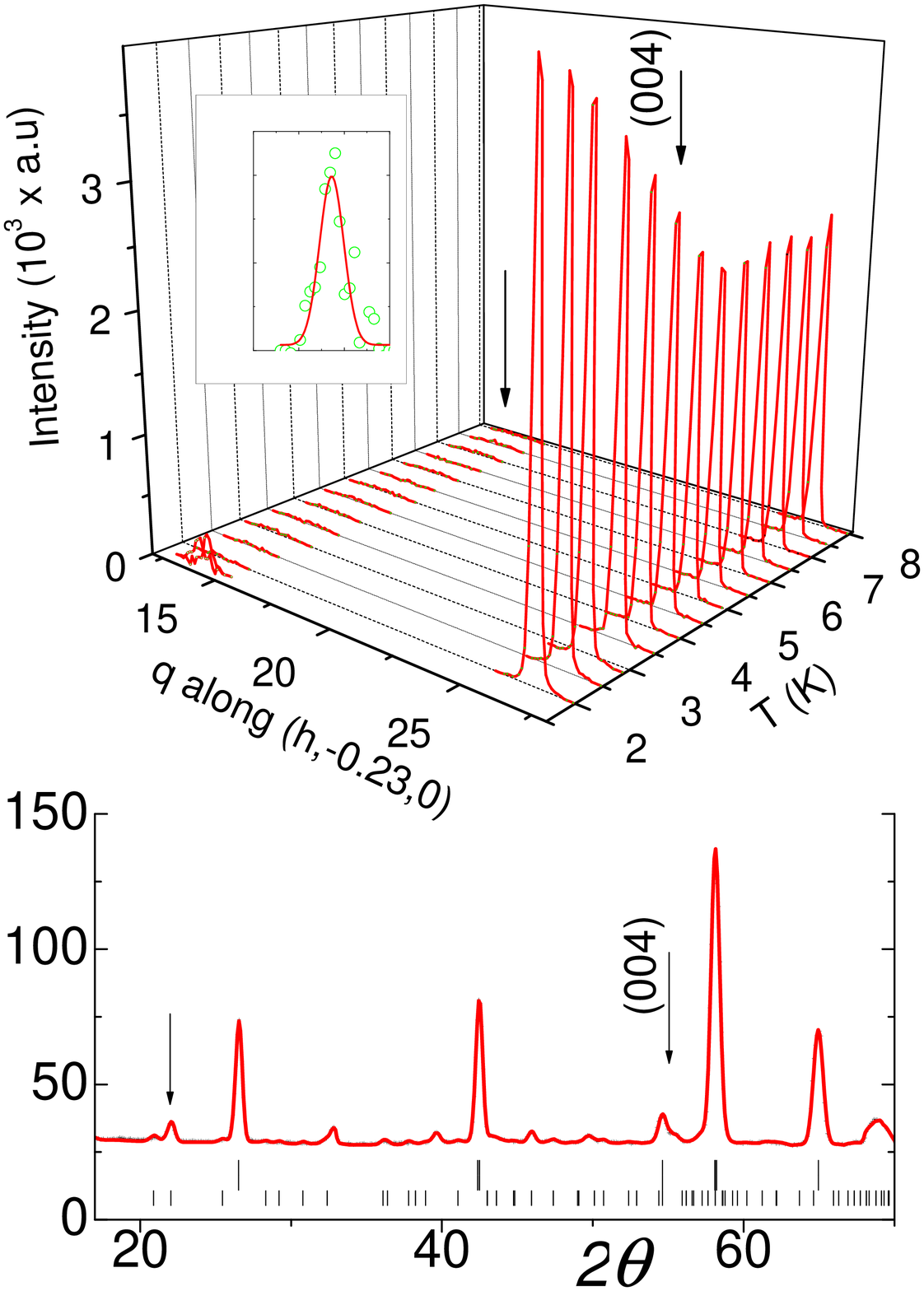}%
\caption{The upper panel shows, on a three dimensional plot, the thermal
evolution of the (0,0,4) peak and an impurity peak. The latter together with
its Gaussian fit is shown, on an expanded scale, in the inset. The lower panel
shows the powder diffractogram at 2 K (taken from Fig.7). The \ intensity of
the (0,0,4) and impurity peaks (marked by a vertical arrow) are almost equal
in the powder diffractrogram but are a factor of 20 different in the
single-crystal diffractograms (no correction for multiplicity factors are
considered). The wave length used for the measurement in the upper panel is
2.3606 $\operatorname{\mathring{A}}$ while for the lower panel is 2.42
$\operatorname{\mathring{A}}.$}%
\label{FM-Fig.9}%
\end{center}
\end{figure}
%EndExpansion

Various $q$-scans within the range 1.7$<T<$8 K were performed on a single
crystal of \textrm{TbCo}$_{2}$\textrm{B}$_{2}$\textrm{C.} A wide range of $q$
space were scanned while maintaining the temperature constant at 1.7 K.
Indeed, most of the nuclear (and ferromagnetic) peaks that satisfy the
relation $h+k+l=2n$ were observed. In addition, we also looked, at 1.7 K, for
any modulated mode within a wide range covering $-0.4<h<0.6$, $-0.4<k<0.4$,
and $-0.3<l<0.3$. Some weak reflections were observed. Fig. \ref{FM-Fig.9}
shows the thermal evolution of one of these peaks and, for comparison, also
that of \ the (0,0,4) peak. As evident, the intensity of the (0,0,4) peak,
being due to magnetic and nuclear contributions, decrease smoothly and goes to
the value of the nuclear intensity as the temperature reaches $T_{c}$. In
particular there is no visible variation in the (0,0,4)\ intensity when $T$ is
varied across $T_{m}$, indicating that the event at this temperature, does not
belong to the main phase. On the other hand, the intensity of the other peak
decays very fast as the temperature increases and is almost within the
experimental uncertainty when $T>$ 2 K. Such a thermal evolution is similar to
the features observed in the magnetization hysteresis (Fig. \ref{FM-Fig.1}),
in the specific heat anomaly (inset of Fig. \ref{FM-Fig.3}), and in the powder
neutron diffractograms (Figs. \ref{FM-Fig.5}-\ref{FM-Fig.7}). This suggests
that this, as well as the other weak peaks, are related to the same impurity
phase as discussed above. Evidently, the ratio of the intensity of this
contaminating peak to that of the (0,0,4) peak is extremely small in the
single-crystal sample; in terms of the above mentioned impurity scenario, this
means that the impurity concentration in the single-crystal sample is much
smaller than the one in the polycrystalline case: this conclusion is supported
by the observation that the intensity of the specific heat event at $T_{m}$ is
hardly evident in the single-crystal sample (see the inset of Fig.
\ref{FM-Fig.3}).

Based on the above conclusions, the powder diffractograms below $T_{\text{m}}$
were analyzed, with the Rietveld method, as a superposition of three patterns:
the nuclear, the FM mode, and a third unidentified contaminating magnetic
phase. The Rietveld analysis of the nuclear and magnetic patterns of the main
phase below $T_{m}$ is straightforward and gave a satisfactorily fit; the
diffractograms are shown in Fig. \ref{FM-Fig.7} while the obtained structural
and magnetic parameters are given in Fig. \ref{FM-Fig.8}; the latter figure
reveal that the lattice parameters evolve smoothly across $T_{m}$ and,
furthermore, the mismatch parameter $(a-b)/a$ is the same as the one
manifested in the magnetostriction experiment of Fig. \ref{FM-Fig.4} (a). On
the other hand, Fig. \ref{FM-Fig.8} (c) compares the reduced Tb magnetic
moment, $\mu(T)/\mu($2 K$)$, with the calculated Brillouin function,
\textit{B}$_{6}$($x$). The total angular momentum\textit{ }quantum number is
taken to be 6, representing that of a free Tb$^{3+}$ ion. Since, as shown
above, there are considerable CEF effects, then this \textit{B}$_{6}$($x$)
curve should be taken as a lower bound; nonetheless, the overall thermal
evolution of $\mu(T)/\mu($2 K$)$ follows reasonably well this \textit{B}$_{6}%
$($x$) curve; in particular, it reveals a smooth and monotonic evolution
across the $T_{\text{m}}$ region. Finally, the intensity contribution of the
unidentified phase was calculated using the so-called profile
matching\cite{Rodriguez-Carvajal93-Profile-Matching} (or pattern
decomposition) procedure. Since the crystal structure parameters of this
unidentified phase are unknown, no significance should be attached to the fit
of the impurity phase, only that all the additional weak peaks are associated
with the impurity phase and that the presence of this phase would not modify
the conclusions reached about the magnetic properties of the \textrm{TbCo}%
$_{2}$\textrm{B}$_{2}$\textrm{C} phase.

\section{Discussion and Conclusions}

The above results show that the paramagnetic susceptibilities are given by the
CW behavior of the Tb$^{3+}$ moments, that this contribution is strong enough
to mask any magnetic contribution of Co-sublattice (if there is any), that
$\mu_{\text{eff}}$\ and $\mu_{\text{sat}}$ are typical of Tb$^{3+}$ ion and
are almost equal to the ones observed in \textrm{TbNi}$_{2}$\textrm{B}$_{2}%
$\textrm{C} (see Table \ref{Table II}), and that the transition at
$T_{\text{c}}$ is identified as being due to the FM order of the Tb
sublattice. In addition, the sample-dependent $T_{\text{m}}$-event is
associated with magnetic contamination. Below we give a further argument in
support of this latter identification. Obviously, if such a $T_{\text{m}}%
$-event is intrinsic, then it must be either due to the Co- or Tb- sublattice.
That the $M(H,T<T_{\text{C}})$ and $^{100}\lambda_{100}(H,T<T_{\text{C}})$
isotherms do not show any metamagnetic transition and that the same FM state
of Tb sublattice is being maintained across $T_{\text{m}}$, then this
$T_{\text{m}}$-event can not be related to a rearrangement (in direction or
strength) of the Tb FM sublattice.

Let us now discuss the claim that the Co orbitals which, being on the verge of
ferromagnetism, are spontaneously polarized. But the fulfilment of this
possibility requires that $H_{\text{eff}}^{\text{Co}}$ $\geq$ $H_{\text{cr}%
}^{\text{Co}}$, a relation which can, \textit{apriori}, be satisfied since
$H_{\text{eff}}^{\text{Co}}$ increases monotonically as $T$ decreases [see
Fig. \ref{FM-Fig.8} (c)]. In this case, there should be two magnetic
transitions: one related to the $R$-subsystem and another to the Co-subsystem;
just as in the case of, e.g., \textrm{Er}$_{\text{0.6}}$\textrm{Y}%
$_{\text{0.4}}$\textrm{Co}$_{\text{2}}$\textrm{.}%
\cite{hauser00-two-peak-(ErY)Co2} Then, as observed in other
intermetallics,\cite{Bloch-Lemaire70-RCo2,Bloch75-RCo2,Cyrot79-electr-cal,Cyrot79-b-electr-cal}
such induced Co-moments should couple ferrimagnetically with the Tb$^{3+}$ FM
sublattice. As there are no spontaneous or field-induced metamagnetic
transitions in the $M(H,T<T_{\text{C}})$ curves, then the possibility of a
spontaneously polarized Co moment must be excluded; as such $H_{\text{eff}%
}^{\text{Co}}$ $<H_{\text{cr}}^{\text{Co}}$( see \S \ III.A).

\begin{table}[th]
\caption{A comparison of the lattice and magnetic parameters of the
isomorphous TbCo$_{2}$B$_{2}$C and TbNi$_{2}$B$_{2}$C. The lattice parameters
are reported for samples at LHe temperatures except the $c$ parameter
of\ TbNi$_{2}$B$_{2}$C; the later was estimated by normalizing its room
temperature value using the values of HoNi$_{2}$B$_{2}$%
C.\cite{Lynn97-RNi2B2C-ND-mag-crys-structure}}%
\label{Table II}%
$%
\begin{tabular}
[c]{cccccccc}\hline\hline
& $a$($\operatorname{\mathring{A}})$ & $b$($\operatorname{\mathring{A}})$ &
$c$($\operatorname{\mathring{A}})$ & $\mu_{eff}$($\mu_{\text{B}})$ &
$T_{C,N}($K$)$ & Magnetic Mode & $\mu_{sp}$($\mu_{\text{B}})$\\\hline
TbCo$_{2}$B$_{2}$C & 3.535 & 3.523 & 10.560 & 9.7 & 6.3 & FM $q$=(000) & 7.6\\
TbNi$_{2}$B$_{2}$C & 3.554$^{a}$ & 3.534$^{a}$ & 10.44$^{b}$ & 9.8$^{c}$ &
15$^{c}$ & LSW, $q$=(0.45,0,0)$^{b}$ & 7.78$^{b,c}$\\\hline\hline
\multicolumn{2}{l}{$^{a}$ Ref. \cite{Song99-Tb-magnetostriction},} &
\multicolumn{3}{c}{$^{b}$ \ Ref.\cite{Lynn97-RNi2B2C-ND-mag-crys-structure},}
& \multicolumn{3}{c}{$^{c}$ \ Ref.\cite{Cho96-TbNi2B2C-anistropy-WF}}%
\end{tabular}
\ \medskip$\end{table}

It is significant that the FM structure of \textrm{TbCo}$_{2}$\textrm{B}$_{2}%
$\textrm{C} is drastically different from any of the reported magnetic
structures of borocarbides,\cite{Lynn97-RNi2B2C-ND-mag-crys-structure} in
particular \textrm{TbNi}$_{\text{2}}$\textrm{B}$_{\text{2}}$\textrm{C} even
though these Tb-based isomorphs are similar in most (if not all) of the
single-ion CEF-influenced properties such as the anisotropy, the strength, and
the orientation of Tb moments (see \S \ III.A and D). However, these isomorphs
are distinctly different (see Table \ref{Table II}) in the value of their
transition points, in their magnetic structures, and in the overall features
of their $H-T$ phase diagrams (a simple one-boundary FM phase versus a cascade
of field-induced phase transitions). These differences suggest that the
effective Tb-Tb magnetic couplings must be different and, furthermore, those
nesting features\cite{Lee94-electronic-structure} that are responsible for the
modulated mode in \textrm{TbNi}$_{\text{2}}$\textrm{B}$_{\text{2}}$\textrm{C}
must be absent in \textrm{TbCo}$_{2}$\textrm{B}$_{2}$\textrm{C}. These
differences, prompted by the introduction of the Co atoms, suggest that the
configuration of their electronic structures [in particular the position of
$E_{F}$\ within the $N$($E$)\ curve and the generalized susceptibilities] must
be different. As mentioned in \S \ I, these arguments are consistent with the
findings of the electronic structure calculation on \textrm{LuCo}$_{2}%
$\textrm{B}$_{2}$\textrm{C}:\cite{Coehoorn94-RNi2B2C-electronic-structure}
these calculations provide an explanation for the surge of the enhanced
paramagnetic character of \textrm{YCo}$_{2}$\textrm{B}$_{2}$\textrm{C} and the
absence of superconductivity in any of the $R$\textrm{Co}$_{\text{2}}%
$\textrm{B}$_{\text{2}}$\textrm{C} compounds (even though \textrm{YCo}$_{2}%
$\textrm{B}$_{2}$\textrm{C} has the same Debye and Sommerfeld coefficients
as\ \textrm{YNi}$_{2}$\textrm{B}$_{2}$\textrm{C}%
).\cite{00-RCo2B2C,04-Pr(CoNi)2B2C}

Finally, it is recalled that the indirect exchange coupling in metallic
magnets are usually written as:\cite{Coqblin-book}%
\[
J(R_{ni})=\frac{9\pi n^{2}\Gamma^{2}(g-1)^{2}}{8V^{2}E_{F}}F(2k_{F}R_{ni}%
)\exp(-R_{ni}/\lambda),
\]%
\[
\text{ }F(x)=\left[  \frac{x\cos(x)-\sin(x)}{(x)^{4}}\right]  ,
\]
where $\Gamma$ is the $s-f$ exchange coupling, $n$ is the carrier
concentration [governed by by $N(E_{F})$], $R_{ni}$ is the distance separating
the moments, $E_{F}$ and $k_{F}$ are the Fermi energy and wave vector, and
$\lambda$ ($>R$) is the mean free path. Considering that these couplings
manifest a quadratic dependence on $N(E_{F})$ and a sensitive sinusoidal
dependence on the moments separating distances, then it is no surprise that
the combination of difference in the electronic
structure\cite{Coehoorn94-RNi2B2C-electronic-structure} and in the lattice
parameters (see Table \ref{Table I}) would lead to strong variation in the
magnitude and sign of the coupling constants and as such to drastic difference
in the magnetic structures of these Tb-based isomorphs. In fact this
difference is not limited to these Tb-based isomorphs, our preliminary studies
on the magnetic structures of\ the $R$\textrm{Co}$_{2}$\textrm{B}$_{2}%
$\textrm{C} series showed that this is valid for the whole $R$\textrm{Co}%
$_{2}$\textrm{B}$_{2}$\textrm{C }magnets:\cite{MagneticStructure-RCo2B2C-2008}
as an example, the FM mode is observed in \textrm{TmCo}$_{2}$\textrm{B}$_{2}%
$\textrm{C} and \textrm{HoC}$_{2}$\textrm{B}$_{2}$\textrm{C}.

\begin{acknowledgments}
We acknowledge the partial financial support from the Brazilian agencies CNPq
(485058/2006-5) and Faperj (E-26/171.343/2005).
\end{acknowledgments}

\bibliographystyle{apsrev}
\bibliography{Borocarbides,Crystalograph,Intermetallic,Mag-classic,Massalami,ND-RepAnalysis,Nuclear-hyperfine-MES,To-Be-Published}

\appendix{}

\section{Magnon Contribution from the ferromagnetic Tb-sublattice of
\textrm{TbCo}$_{2}$\textrm{B}$_{2}$\textrm{C}}

The above-mentioned ferromagnetic order of the Tb-sublattice suggests that the
dominant exchange interactions within the same layer (approximated by a
positive $J_{1}$) as well as those among different layers (approximated by a
positive $J_{2})$ are ferromagnetic. Let us assume that the main contributions
to the magnetic energy is due to the above-mentioned exchange couplings and
anisotropic crystalline electric field interactions. Within the low
temperature regime of interest, the anisotropic interactions can be
approximated by an effective field $\vec{H}_{a}$ (in energy units) which, for
this particular\ case, forces the moments to points along the $a$ axis. The
Hamiltonian (under zero external field) can be written as:
\begin{equation}
\mathcal{H}=-\sum_{i,j\epsilon A,B}J_{1}\vec{S_{i}}\cdot\vec{S_{j}}%
-\sum_{<ij>,i\in A,j\in B}J_{2}\vec{S_{i}}\cdot\vec{S_{j}}-\vec{H_{a}}%
.\sum_{i\in A,B}\overrightarrow{S_{i}} \label{H}%
\end{equation}
All symbols have their usual meanings. The first term sums the bilinear
products of two neighboring spins of the the same layer $A$ and afterwards the
contribution of all layers are added together. The second term sums all the
bilinear product of two neighboring spins (each belong to a different but an
adjacent layer). The third term is a sum over all single-ion anisotropic
energies. Using standard linear spin-wave approximation (considering a
non-interacting magnon gas), Eq. \ref{H} can be diagonalize to give the
following dispersion relation:
\begin{align*}
\hbar\omega_{k}  &  =SJ_{1}(4-2\cos bk_{x}-2\cos ak_{z})+\\
&  2SJ_{2}\left\{
\begin{array}
[c]{c}%
4-\cos(\frac{a}{2}k_{z}+\frac{b}{2}k_{x}+\frac{c}{2}k_{y})-\cos(-\frac{a}%
{2}k_{z}+\frac{b}{2}k_{x}+\frac{c}{2}k_{y})-\\
-\cos(\frac{a}{2}k_{z}-\frac{b}{2}k_{x}+\frac{c}{2}k_{y})-\cos(\frac{a}%
{2}k_{z}+\frac{b}{2}k_{x}-\frac{c}{2}k_{y})
\end{array}
\right\}  +H_{a},
\end{align*}
where $z$ (the quantization axis), $x$, $y$ axes are, respectively, along the
$a\ $(easy direction), $b$, $c$ directions of the crystallographic unit cell.
In the long wave limit, up to second order, this simplifies to:
\begin{equation}
\hbar\omega_{k}=\Delta+c_{x}k_{x}^{2}+c_{y}k_{y}^{2}+c_{z}k_{z}^{2}.
\label{w-long-limit}%
\end{equation}

Assuming a weaker orthorhombic distortion (a$\approx$b and $c_{x}\approx
c_{z}$), one gets:
\begin{align*}
c_{x}  &  =Sa^{2}(J_{1}+J_{2})\equiv c_{z}\\
c_{y}  &  =Sc^{2}J_{2}.
\end{align*}

The energy gap ($k=0$) is:%
\begin{equation}
\Delta=H_{a} \label{Delta}%
\end{equation}

The expression for the density of states (obtained from integrating over the
constant energy surface, $\epsilon\equiv\omega$) is:
\begin{equation}
\rho(\epsilon)={\frac{V}{(2\pi)^{3}}}\int{\frac{dS_{\epsilon}}{\mid\nabla
\hbar\omega_{k}\mid}}={\frac{V}{2\pi^{2}}}{\frac{\sqrt{\epsilon-\Delta}}%
{c_{x}\sqrt{c_{y}}a^{2}c}}\text{.} \label{ro}%
\end{equation}

The magnon contribution to the total energy is:
\[
E=E_{0}+\int_{\Delta}^{\infty}d\epsilon{\frac{\epsilon\rho(\epsilon)}%
{e^{\beta\epsilon}-1}}%
\]
where $E_{0}$ is a constant independent of temperature. Using Eqs.
\ref{w-long-limit}, \ref{Delta}, and \ref{ro}, the molar specific heat is
($\nu$ is the number of moles):
\[
C_{mag}(T)={\frac{1}{\nu}\frac{dE}{dT}}={\frac{R}{2\pi^{2}c_{x}\sqrt{c_{y}}}%
}{\frac{1}{T^{2}}}\sum_{n=1}^{\infty}{\ I_{n},}%
\]
where
\[
I_{n}=n\int_{\Delta}^{\infty}d\epsilon\epsilon^{2}\sqrt{\epsilon-\Delta
}e^{-n\beta\epsilon}=2ne^{-n\beta\Delta}\{\frac{1.3.5.\sqrt{\pi}}{2^{4}%
(n\beta)^{7/2}}+2\Delta\frac{1.3.\sqrt{\pi}}{2^{3}(n\beta)^{5/2}}+\Delta
^{2}\frac{\sqrt{\pi}}{2^{2}(n\beta)^{3/2}}\}.
\]
The final result can be rearranged to give:
\begin{equation}
C_{mag}(T)=\frac{15.R\Delta^{\frac{3}{2}}}{8.\pi^{\frac{3}{2}}D^{\frac{3}{2}}%
}\sum_{n=1}^{\infty}n.\exp(-\frac{n\Delta}{T})\left[  \frac{4}{15}\left(
\frac{T}{n\Delta}\right)  ^{-\frac{1}{2}}+\left(  \frac{T}{n\Delta}\right)
^{\frac{1}{2}}+\left(  \frac{T}{n\Delta}\right)  ^{\frac{3}{2}}\right]
\label{Cm-FM}%
\end{equation}
where
\begin{equation}
D=2^{\frac{1}{3}}S(J_{1}+J_{2})^{\frac{2}{3}}J_{2}^{\frac{1}{3}} \label{D}%
\end{equation}
is the spin-wave stiffness coefficient\cite{Kittel-SpinWave} which is a
measure of the effective coupling strength. This expression reproduces the
high temperature limit $(T>>\Delta)$,
\[
C_{M}(T)\propto e^{-\Delta/T}T^{3/2},
\]
which for $\Delta=0$ gives the well known $\frac{3}{2}$-Bloch
expression.\cite{Kittel-SpinWave} At lower temperatures $(T<<\Delta)$, the
strong dependence on the gap is emphasized by the expression:
\[
C_{M}(T)\propto e^{-\Delta/T}T^{-1/2}.
\]

\end{document}